\documentclass[
    ,final            
  ,numberedheadings 
  ]
  {aipproc}

\layoutstyle{8x11single}

\begin{document}

\title{Everything Under the Sun: A Review of Solar Neutrinos}


\author{Gabriel D. Orebi Gann}{
  address={Lawrence Berkeley National Laboratory, Berkeley, CA 94720, USA \\ Department of Physics, University of California, Berkeley, CA 94720, USA}
}


\begin{abstract}
 
Solar neutrinos offer a unique opportunity to study the interaction of neutrinos with matter, a sensitive search for potential new physics effects, and a probe of solar structure and solar system formation.  This paper describes the broad physics program addressed by solar neutrino studies, presents the current suite of experiments programs, and describes several potential future detectors that could address the open questions in this field.  This paper is a summary of a talk presented at the Neutrino 2014 conference in Boston.

 \end{abstract}

\maketitle


\section{Introduction}
\begin{figure}[!h]
\includegraphics[width=0.65\textwidth]{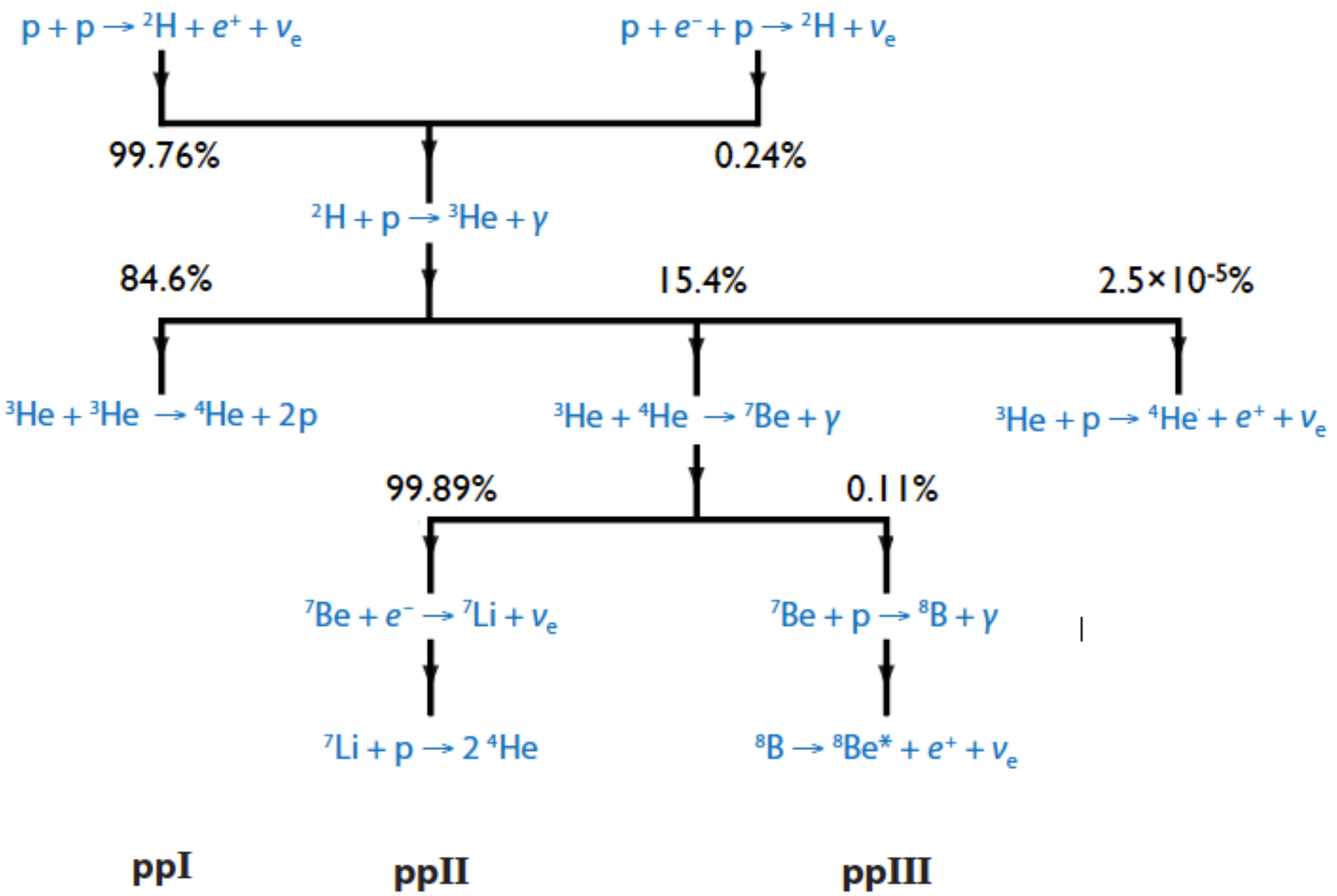}
\includegraphics[width=0.34\textwidth]{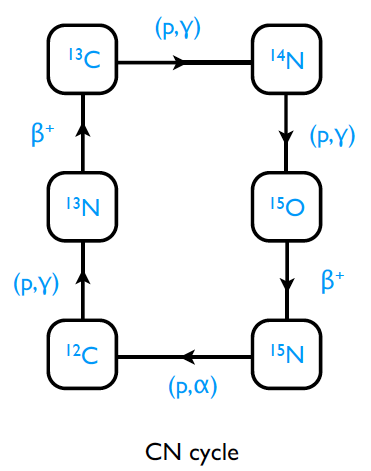}
\caption{ Neutrino production in solar fusion reactions showing: (Left) The three principal cycles in the pp chain and (Right) The CN I cycle, which produces the $^{13}$N and $^{15}$O neutrinos.  Taken from~\cite{HRS}
\label{f:fusion}}
\end{figure}

Solar neutrinos are produced as a by-product of the nuclear fusion reactions that power the Sun.  Neutrinos are produced in two fusion cycles, as shown in Fig.~\ref{f:fusion}: the pp chain, the primary source of solar power, and the sub-dominant CNO cycle~\cite{HRS}.  In all cases the neutrinos are produced in the electron flavor state.  

\begin{figure}[!h]
\includegraphics[width=0.5\textwidth]{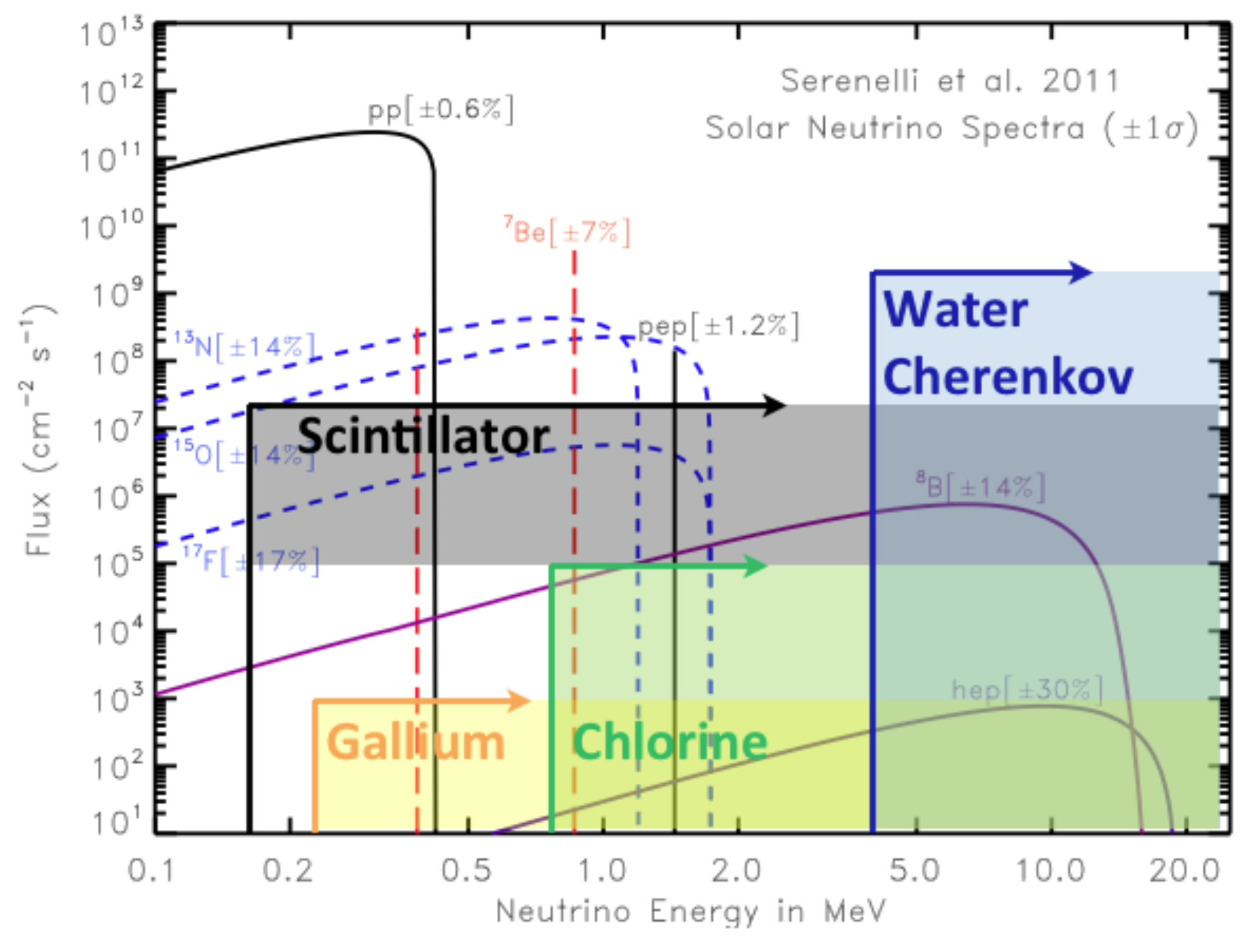}
\caption{Spectra of neutrinos emitted by fusion reactions in the Sun.  Solid lines represent neutrinos from the pp chain and dashed lines are neutrinos from the CNO cycle.  Original image taken from ~\cite{HRS}, and modified by the author to include the sensitivity of various experimental approaches.
\label{f:spec}}
\end{figure}

\begin{figure}[!h]
\includegraphics[width=0.8\textwidth]{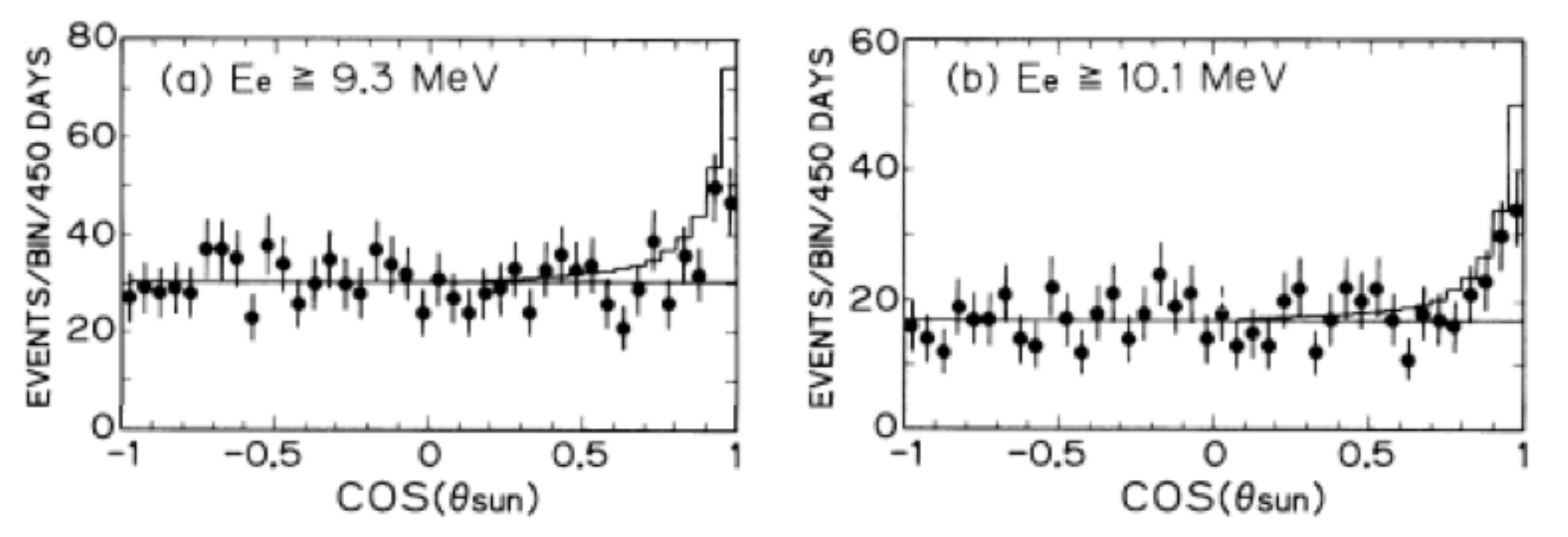}
\caption{ Distribution of the cosine of the angle between the reconstructed trajectory of an electron and the direction of the Sun at a given time, from Kamiokande-II~\cite{kii}.
\label{f:Kii}}
\end{figure}

Neutrinos produced in these different cycles can be distinguished by their energy spectra.  Figure~\ref{f:spec} shows the spectrum of neutrinos emitted in each process: solid lines represent neutrinos from the pp chain, and dashed lines are neutrinos from the CNO cycle.  Also shown are the energy regimes in which these neutrinos have been detected.  
The first experiment to detect neutrinos from the Sun was the Chlorine experiment of Ray Davis {\emph et al.} at the Homestake mine in South Dakota~\cite{chlorine}.  These observations were supported by later measurements from gallium-based experiments:   GALLEX~\cite{gallex}; SAGE~\cite{sage}; and GNO~\cite{gno}. 
These radio-chemical experiments can achieve very low energy thresholds, but perform an integral measurement of all neutrinos above threshold, producing a single integrated flux measurement.  
Water Cherenkov experiments such as Super-Kamiokande~\cite{sk} and the Sudbury Neutrino Observatory~\cite{snocc} have higher thresholds, but can perform real-time detection thus allowing extraction of both directional and spectral information.  This capability allowed Kamiokande-II to first demonstrate that the observed neutrinos were in fact coming from the Sun (Fig.~\ref{f:Kii}).  
Scintillator experiments have recently demonstrated thresholds competitive with the radio-chemical approach~\cite{Borexpp}, and can also perform real-time detection, making this the lead candidate technology for future detectors.

For several decades there was a large discrepancy between the flux of solar neutrinos predicted by the Standard Solar Model (SSM)~\cite{ssm} and that measured by terrestrial experiments.  This became known as the Solar Neutrino Problem.  
The Sudbury Neutrino Observatory (SNO) experiment resolved the Solar Neutrino Problem with measurements in 2001 and 2002 by detecting the Sun's ``missing'' neutrinos, confirming the theory of neutrino flavor change.  SNO's use of a heavy-water target opened up both the charged-current (CC) and neutral-current (NC) interactions for neutrino detection.  At solar neutrino energies the CC interaction is sensitive only to electron neutrinos, whereas NC is sensitive equally to all three flavors.  
The combination of SNO's CC measurement with Super-Kamiokande's high-precision elastic scattering (ES) measurement demonstrated that the electron neutrinos produced in the Sun were oscillating to other flavors prior to detection~\cite{snocc}, a result later confirmed at 5$\sigma$ by SNO's measurement of the flavor-independent $^8$B flux using the NC interaction~\cite{snonc}.  This flavor change was confirmed as being due to oscillation soon thereafter by the KamLAND experiment's observation of the oscillation pattern in the spectrum of antineutrinos emitted by nearby nuclear reactors~\cite{kl}. 
This opened the door to a precision regime, allowing neutrinos to be used to probe the structure of the Sun, as well as the Earth and far-distant stars.

\section{Physics Beyond the Solar Neutrino Problem}

With the resolution of the Solar Neutrino Problem, this field now offers a rich and diverse program of physics, including  understanding neutrino properties and behavior, and using neutrinos as a probe to better understand solar structure and solar system formation. 
Solar neutrinos offer the only observed effect of the interaction with matter on neutrino propagation to date, making this a unique opportunity to study and further understand this effect.  

\subsection{The Search for New Physics}
The phenomenon of solar neutrino oscillation is complicated by the interactions of neutrinos with matter as they propagate outwards from their production point.  The three flavors interact differently with matter: electron neutrinos can interact via CC as well as NC interactions, causing an additional diagonal term in the Hamiltonian in matter, which increases the effective mass of the electron neutrino.  This ``MSW'' effect, as proposed by Mikheyev, Smirnov and Wolfenstein~\cite{wolf, ms}, enhances oscillation in an energy-dependent fashion.  As a result, neutrinos produced in different fusion reactions are affected to a different extent, due to their different energies.  The effect is negligible at sub-MeV energies, where vacuum oscillation dominates, resulting in a electron neutrino survival probability of a little over one half.  This is supported by observations from the gallium experiments GALLEX~\cite{gallex}, SAGE~\cite{sage}, and GNO~\cite{gno}, which were primarily sensitive to the low-energy pp neutrinos.  
 Matter effects become significant above about 5~MeV, with the additional suppression of electron neutrinos resulting in a survival probability of approximately one third, as seen by water Cherenkov experiments such as Super-Kamiokande~\cite{sk} and SNO~\cite{snocc}. 
A transition is predicted in between these two regimes, where the survival probability falls from the vacuum-averaged value to the additionally-suppressed matter oscillation value.  
In both the vacuum- and matter-dominated regions the survival probability is determined by the value of the mixing angle, $\theta_{12}$, and not by the details of the interaction of neutrinos with matter.  In order to probe this interaction, one must probe the vacuum-matter transition region, in an energy range of $1-5$~MeV.  

\begin{figure}[!h]
\includegraphics[width=0.28\textwidth]{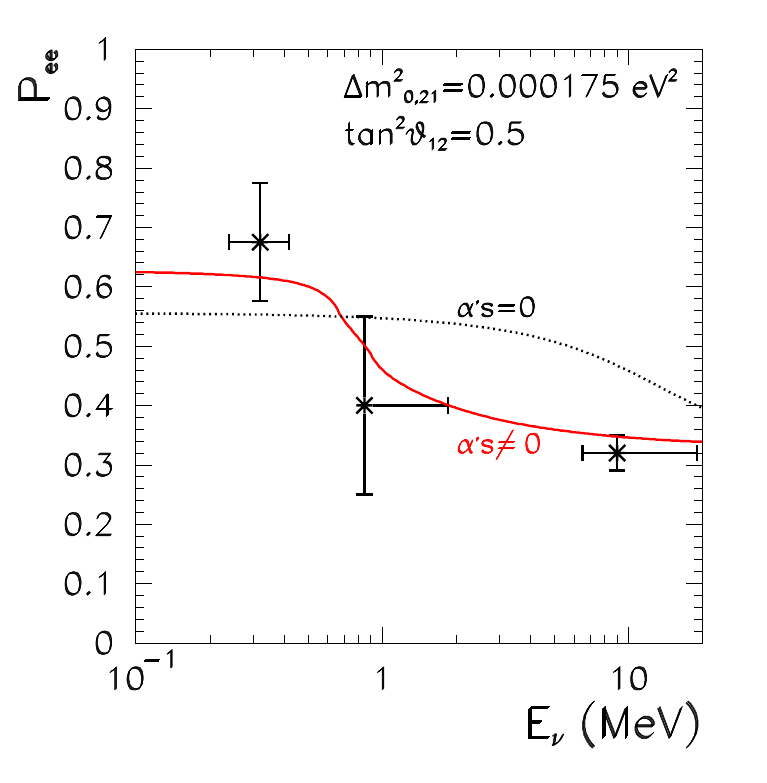}
\includegraphics[width=0.35\textwidth]{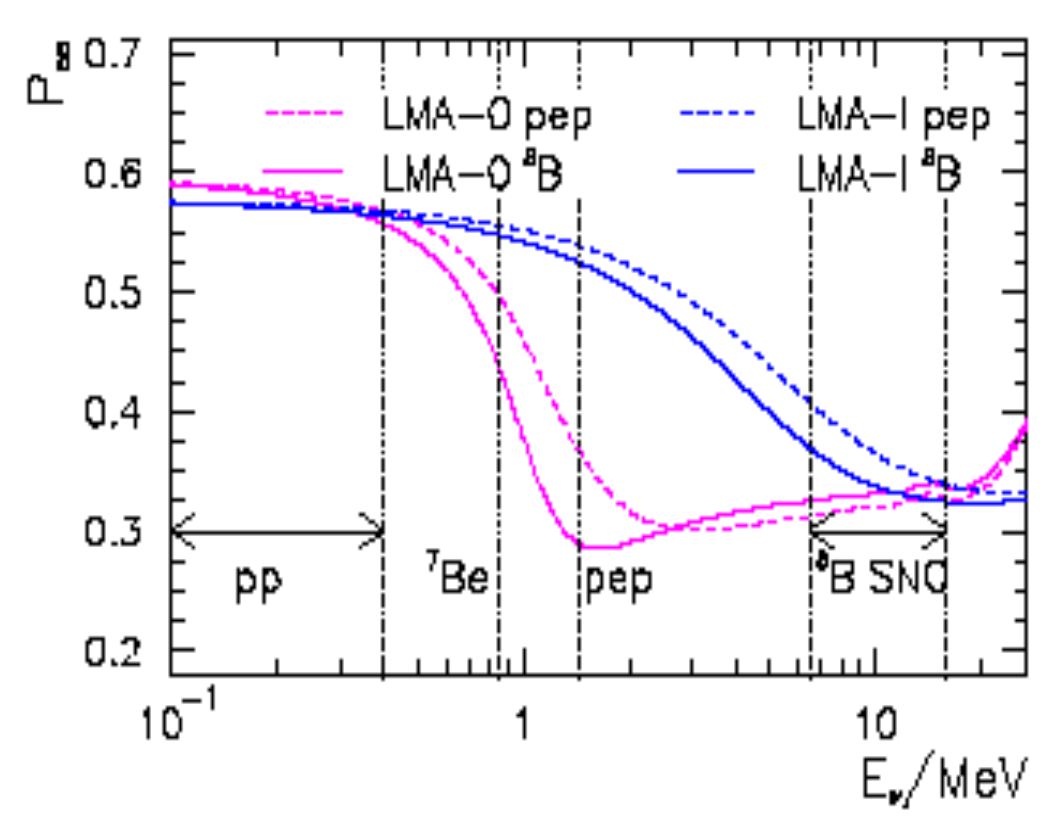}
\includegraphics[width=0.36\textwidth]{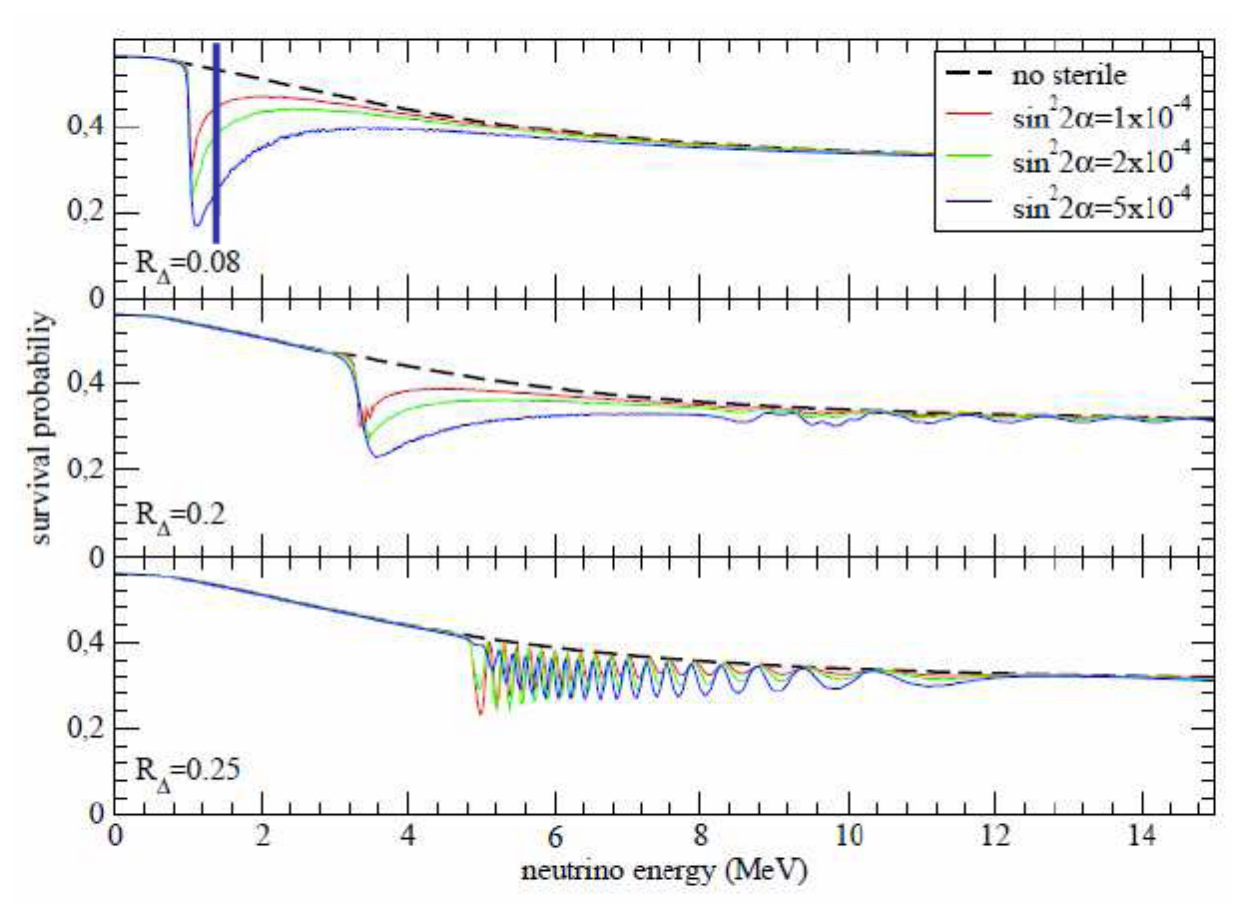}
\caption{ The effect of several new physics phenomena on the vacuum-matter transition region in the solar electron neutrino survival probability ($P_{ee}$).  From left to right: mass-varying neutrinos~\cite{mavan}; non-standard interactions~\cite{nsi2}; and sterile neutrinos~\cite{sterile}.
\label{f:pee}}
\end{figure}

A sensitive probe of this region is not only required to test our understanding of the MSW effect, it is also extremely sensitive to potential new physics effects.  Non-standard interactions such as flavor-changing neutral currents, sterile neutrinos, or mass-varying neutrinos can alter the shape and position of the so-called ``MSW rise'', as illustrated in Figure~\ref{f:pee}.

Several experiments have sought to probe this region, led by SNO's low-threshold analysis~\cite{leta}, which also performed the first direct fit for the energy-dependent survival probability.  Super-Kamiokande have recently performed a similar analysis~\cite{skleta}, and several experiments have extracted the spectrum of recoil electrons from the ES interaction, including Borexino~\cite{Bor8B}, Super-Kamiokande~\cite{SK8B}, and KamLAND~\cite{KL8B}.  In no case is there clear evidence for the predicted rise in the spectrum at low energies.  In fact, in several cases there is some evidence for a weak turn down, most clearly present in the SNO data but also visible in the Borexino and KamLAND spectra when overlaid with theoretical predictions.  This turn down could be evidence for new physics.  

\begin{figure}[!h]
\includegraphics[width=0.4\textwidth]{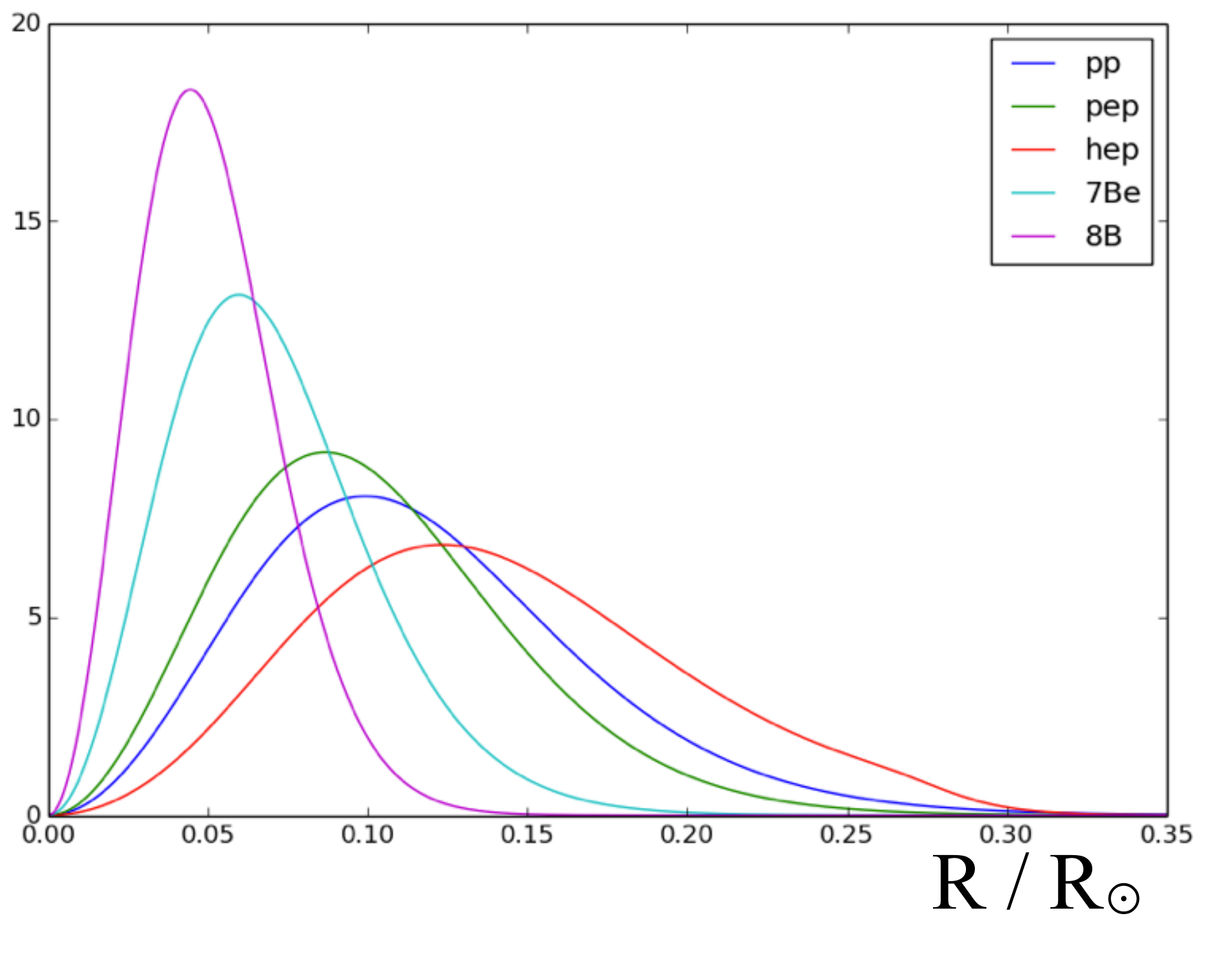}
\includegraphics[width=0.54\textwidth]{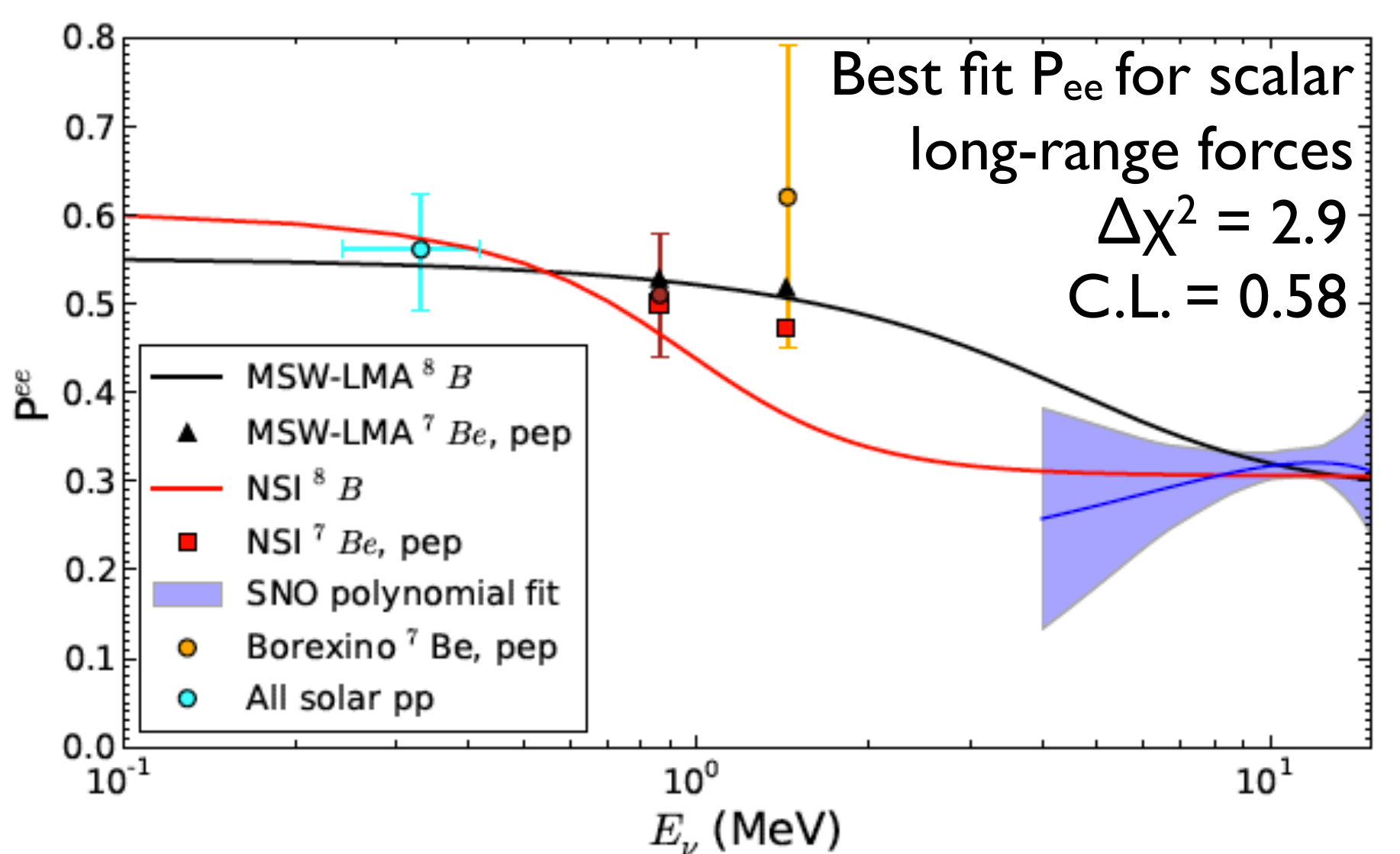}
\caption{(Left) The production region of different neutrino species as a function of solar radius~\cite{richie2}.  (Right) The effect of long-range forces on the electron neutrino survival probability, showing the best fit to the global solar neutrino data set~\cite{richie}.  The effect on the $^8$B neutrino survival probability is large, and could be tightly constrained by additional experimental data, whereas the shift in the pep survival probability is very small and would require percent-level precision to distinguish from standard MSW oscillation.
\label{f:8b}}
\end{figure}

A number of papers have sought to interpret this data in the context of potential new physics effects.  One such paper did a comprehensive study of several effects, including: non-standard forward scattering; mass-varying neutrinos; long-range leptonic forces; and the possibility of a non-standard solar model, such as changes in core density~\cite{richie}.  This paper showed that there are no significant effects (all effects were present at less than two sigma significance), but critically also showed that the results were limited by experimental precision.  Although the pep neutrinos, a line source at 1.44~MeV, lie well into the transition region, they are in fact less sensitive to these effects than the $^8$B neutrinos due to their production further out from the core of the Sun.  This is illustrated in Figure~\ref{f:8b}, which shows the effect on the $^8$B and pep neutrino survival probability in the presence of scalar long-range forces.  
The $^8$B neutrino spectrum is a continuous spectrum with an end point around 16~MeV (Fig.~\ref{f:spec}), thus the $^8$B neutrino survival probability is likewise continuous across the energy range shown.  Both $^7$Be and pep neutrinos are line sources and, thus, the survival probability is only evaluated at that single energy.  The survival probability for $^7$Be and pep neutrinos can differ from that for $^8$B neutrinos at the same energy due to their different production regions in the Sun.  The survival probability for each neutrino species under the standard MSW scenario is shown in black, and the effect of non-standard interactions (NSI) is shown in red.  The effect on the $^8$B neutrino survival probability is large, and could therefore be tightly constrained by additional experimental data, whereas the shift in the pep survival probability is very small and would require percent-level precision to distinguish from standard MSW oscillation.  A precision measurement of the low-energy $^8$B spectrum is therefore called for in order to probe this region with the required sensitivity.

\subsection{Solar Metallicity}

One of the critical factors that engendered confidence in the SSM was the excellent agreement ($\sim0.1\%$) of SSM predictions for the speed of sound with helioseismological measurements.  The speed of sound predicted by the SSM is highly dependent on solar dynamics and opacity, which are affected by the Sun's composition.  In recent years more sophisticated models of the solar atmosphere have been developed, replacing one-dimensional modeling with fully three-dimensional modeling, and including effects such as stratification and inhomogeneities.  These models produce results more consistent with neighboring stars of similar type, and yield improved agreement with absorption-line shapes~\cite{asplund}.
However, they also result in an abundance of metals (elements heavier than H or He) in the photosphere that is $\sim$30\% lower than previous values, which, when input into the SSM, produces a discrepancy with helioseismological observations.  This new disagreement has become known as the ``Solar Metallicity Problem''~\cite{met1, met2}.
Although the impact of these changes on the pp-chain neutrinos is small relative to theoretical uncertainties, the neutrino flux from the sub-dominant CNO cycle depends linearly on the metallicity of the solar core, and the predictions for the two models differ by greater than 30\%.  The theoretical uncertainty on these predictions is roughly 14--18\%, but this can be reduced to $<$10\% using correlations in the theoretical uncertainties between the CNO and $^8$B neutrino fluxes: the two have similar dependence on environmental factors, thus a precision measurement of the $^8$B neutrino flux can be used to constrain the CNO neutrino flux prediction.  A precision measurement of the CNO flux then has the potential to resolve the current uncertainty in heavy element abundance.

A measurement of these neutrinos can also test our understanding of heavier main-sequence stars, in which the CNO cycle dominates over the pp fusion chain.  A key assumption in the SSM is that convective mixing produced a homogeneous Sun, and that subsequent evolution has not appreciably altered the metal distribution.  A measurement of CNO neutrinos could be used to test this postulate.  Further, a more challenging goal would be independent measurements of the  $^{13}$N and $^{15}$O neutrino fluxes (two of the three sources that contribute to the overall CNO flux), which would determine the separate primordial abundances of C and N~\cite{HRS} and allow us to test the extent to which the CNO cycle is in equilibrium inside the Sun.

\subsection{Other Physics}
\paragraph{Day-Night Effect}
In addition to the predicted transition region in the energy spectrum, current understanding of the MSW effect predicts that there should be some regeneration of electron neutrinos as they pass through the Earth.  This would result in an asymmetry in the measured flux between day-time and night-time observations.  The predicted effect is very small and results to date are limited by statistics.  Data from Super-Kamiokande shows a strong hint of this effect at 2.7$\sigma$~\cite{skdn}, but a definitive observation requires an increased global data set.  This is a requirement in order to confirm our understanding of the MSW effect, and the interactions of neutrinos with matter.

\paragraph{Solar Luminosity}
The so-called luminosity constraint makes the assumption that solar neutrinos are produced in the same fusion reactions that power the Sun.  A precision measurement of the total solar neutrino flux would allow a test of this constraint, which would not only test for additional mechanisms for energy loss or generation in the Sun, but also allow us to monitor the Sun's output over recent years using neutrinos.  This test requires a percent-level measurement of the pp solar neutrino flux.  

\paragraph{Flux and Oscillation Parameters}
Precision measurements of neutrino flux and oscillation parameters would allow us to test for symmetries with mixing in the quark sector, which could provide an early step on the road towards Grand Unified Theories.


\section{Experimental Program}
Current and future solar neutrino experiments can be separated into two categories according to their method for neutrino detection: elastic-scattering (ES) or charged-current (CC) detection.  The former lends itself to high statistics.  However, the differential cross section is very broad, limiting the precision with which the neutrino energy spectrum can be extracted from the spectrum of recoil electrons.  CC detection allows a high-precision measurement of the energy spectrum due to the much more strongly peaked cross section.

\subsection{Elastic-Scattering Detection}

ES detectors tend to follow a similar design: a light producing target housed in a large, monolithic detector.  These experiments can be further broken down according to the target material.

\subsubsection{Water Cherenkov}

Super-Kamiokande has collected an impressive solar neutrino data set during its operation, over four phases of data taking.  Recent results include an extraction of the electron neutrino survival probability to low energies~\cite{skleta}, and first evidence for the day-night effect at $2.7\sigma$~\cite{skdn}.  This latter result is still limited by the statistical precision.  The proposed Hyper-Kamiokande project~\cite{hk} would have a target mass 20 times that of Super-Kamiokande, providing unprecedented sensitivity to this effect at the 0.5\% level.  This would allow confirmation of MSW via observation of the day-night effect at $4\sigma$.

\subsubsection{Organic Scintillator}

Liquid scintillator has a much higher light yield than the Cherenkov effect relied upon by water detectors, resulting in roughly 50 times more photons per MeV of deposited particle energy.  As a result these detectors can achieve lower thresholds and improved resolution.  Organic scintillator can also be made extremely clean, providing an ultra-low background environment.

Borexino is a 300 ton detector located in Gran Sasso in Italy.  The Borexino collaboration has achieved unprecedented low levels of contamination in their scintillator: the intrinsic contamination from $^{232}$Th has been reduced to less than $1\times10^{-18}$~g/g, and less than $8\times 10^{-20}$~g/g of $^{238}$U.  As a result Borexino have achieved an impressive suite of solar neutrino results, including the highest precision measurement of $^7$Be neutrinos~\cite{Borex7be}, the tightest constraint (upper bound) on CNO neutrinos to date~\cite{Borexpep}, first evidence of pep neutrinos~\cite{Borexpep}, and, perhaps most impressive, the first direct measurement of pp neutrinos~\cite{Borexpp}.  

SNO+ is a new solar neutrino experiment under construction in the SNOLAB facility in Canada, which will 
re-use the SNO detector, replacing the heavy water target with liquid scintillator.  SNO+ will use 780~ton of linear alkylbenzene (LAB), developed by collaborators at Queen's University, in Canada.  
A potential major source of background is cosmic muons striking organic molecules in the detector, producing unstable isotopes such as $^{11}$C.  The location of SNO+ at $\sim$6000~m.w.e. is a great advantage, since the muon flux drops exponentially with depth: the flux of cosmic muons in SNO+ will be only $\sim$ three per hour.  This cosmogenic background has been a limiting factor for Borexino's measurement of CNO and pep neutrinos, as well as the low-energy $^8$B spectrum.  If SNO+ can achieve the same low levels of radioactive background as Borexino, then the larger detector and low cosmogenic background will allow for better than 10\% (5\%) precision on the pep neutrinos with 1 (3) years of data, a $\sim$15\% measurement of the CNO flux, and an extraction of the ES spectrum of recoil electrons from low-energy $^8$B neutrino interactions.

The LENA experiment is a proposed 50~kton experiment in Europe, with 30\% photocathode coverage.  The timescale for this experiment is currently uncertain.  Should this effort move forwards, LENA would have unprecedented statistics in the low-energy region, providing $3\sigma$ discovery potential for 0.1\%-amplitude temporal modulations in the $^7$Be flux, as well as sensitivity to CNO neutrinos and the low-energy $^8$B spectrum~\cite{lena}.  In addition to ES detection, LENA offers the potential for CC detection on $^{13}$C in the scintillator, increasing the sensitivity to possible spectral distortions.

JUNO is a second-generation reactor antineutrino experiment that is moving ahead in China.  At 700~m rock overburden JUNO is not deep enough for am extensive low-energy program due to the high rate of cosmogenic backgrounds.  However, JUNO will perform a high-precision measurement of many of the oscillation parameters using the reactor antineutrino flux, including a quoted precision of $\sim 0.7$\% on $\sin^2\theta_{12}$~\cite{JUNO}.  This would be a critical input to interpretation of the global solar neutrino data set in the context of potential new physics.

\subsubsection{Inorganic Scintillator}

Inorganic scintillator detectors provide the potential for high-precision measurement of pp neutrinos due to the ultra-low thresholds, and the absence of the dominant background in organic scintillator detectors: $^{14}$C.

Perhaps the most popular choice for the target is liquid neon, which has the potential to provide a background-free fiducial volume.  Neon has the advantage of having no long-lived radioactive isotopes, as well as offering pulse-shape discrimination for rejection of neutron backgrounds.  Operation at 27~K would ensure most contaminants freeze out, and fiducialization can be used to remove game backgrounds.  A 100-ton scale detector, such as the proposed CLEAN experiment~\cite{clean} could achieve percent-level precision on a measurement of the pp solar neutrino flux.

The xenon-based dark matter detectors (LZ~\cite{lz}, XMASS~\cite{xmass}) could also detect pp neutrinos, although a precision measurement would require depletion of $^{136}$Xe by a factor of at least 100 due to the high-rate two-neutrino double beta decay background.

The 34~kton liquid argon TPC detector proposed by the LBNE collaboration would also have some sensitivity to solar neutrinos~\cite{lbne}.  The relatively high threshold ($\sim 5$~MeV) would limit sensitivity to the low-energy spectrum, but the CC interaction on $^{40}$Ar could allow detection of the day-night asymmetry in the high-energy ($> 5$~MeV) region of the spectrum.

\subsection{Charged-Current Detection}

Proposed charged-current detectors use an organic liquid scintillator target loaded with a metallic isotope as a target for CC detection.  There are currently two proposed projects: the segmented LENS detector, and the monolithic ASDC experiment.

\subsubsection{Segmented Detector Design}
\begin{figure}[!h]
\includegraphics[width=0.45\textwidth]{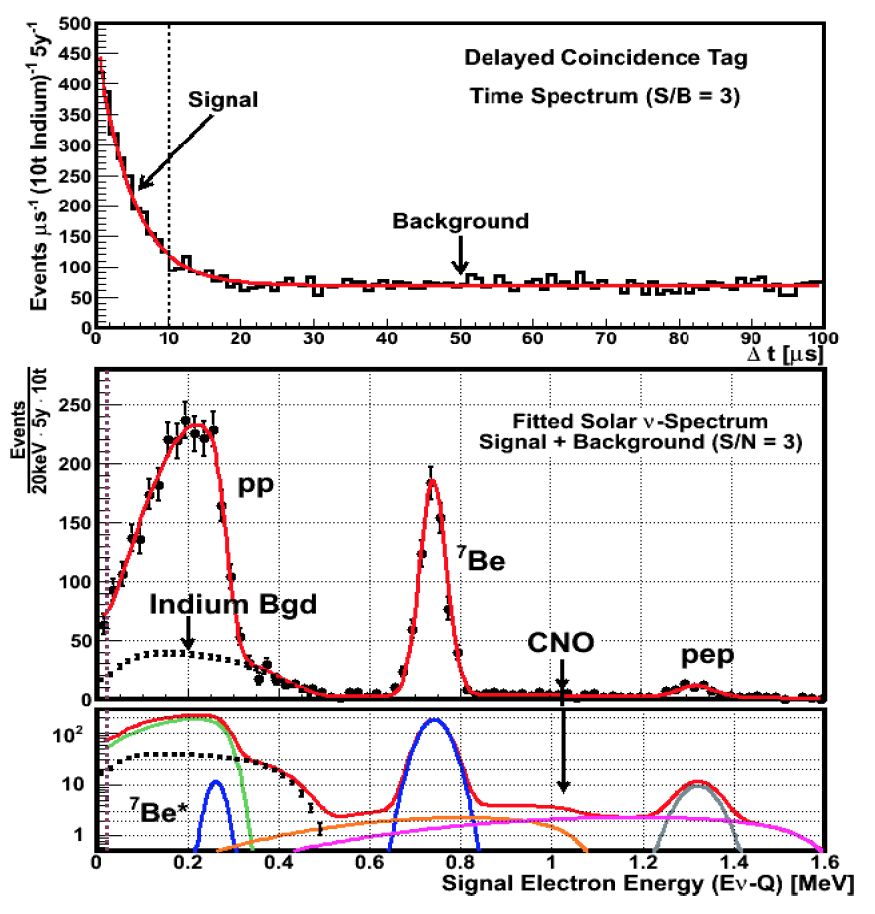}
\caption{Simulated signal spectra from 5 years of data in a 10T-scale LENS detector, taken from~\cite{lens}.  (Top) Delayed coincidence time spectrum fitted to isomeric lifetime $\tau$=4.76 $\mu$s. (Middle)  Energy spectrum($\sim$2000 events in the pp peak) at delays $<10\mu$s, and random coincidences (from the pure $\beta$-decay of In target) at long delays. (Bottom)  Detail of weak CNO signal (in logarithmic scale).
\label{f:lens}}
\end{figure}

The LENS solar neutrino experiment~\cite{lens} proposes to use the CC interaction of neutrinos on $^{115}$In for neutrino detection.  This gives a triple coincidence signal, providing excellent background rejection, and segmentation of the detector assists in rejection of external background events.  The Q-value of the interaction is 115~keV, making the detector sensitive to 95\% of the pp spectrum.  A 10-ton LENS detector could perform a 2.5\%-level measurement of the pp-neutrino flux with 5 years of data (Fig.~\ref{f:lens}).  A CNO measurement is possible but challenging, due to low statistics, and the precision necessary to distinguish between metallicity models would require a larger detector.  A low-energy $^8$B measurement is beyond the reach of this experiment at the proposed scale.

\subsubsection{Monolithic Detector Design}

\begin{figure}[!h]
\includegraphics[width=0.8\textwidth]{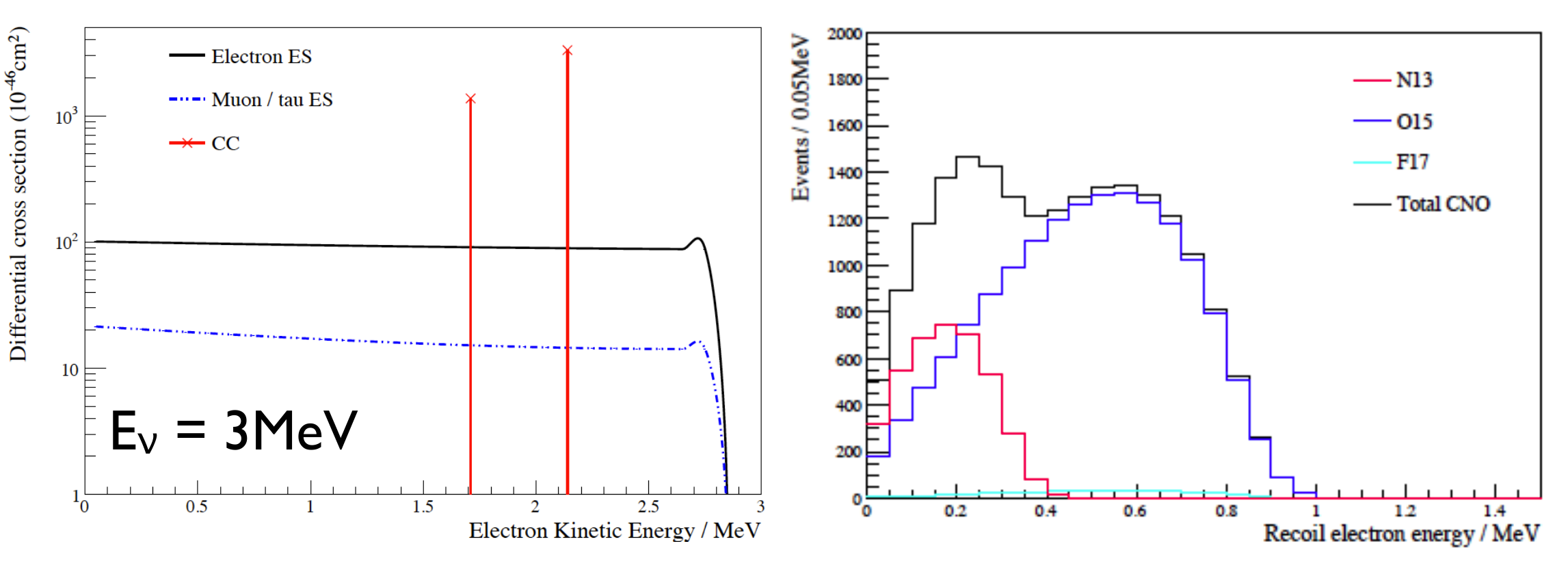}
\caption{(Left) Comparison of the differential cross section for CC on $^7$Li (red) to the ES cross section for electron neutrinos (black) and muon / tau neutrinos (blue dashed) for 3~MeV neutrinos.  The two lines shown for the $^7$Li cross section correspond to the ground and first excited states.  (Right) Predicted spectrum of CNO neutrino events in a 50-kton ASDC detector.
\label{f:asdc}}
\end{figure}

The ASDC (Advanced Scintillation Detector Concept) is a proposed large-scale detector with a water-based liquid scintillator (WbLS) target and high photocathode coverage using ultra-fast, high efficiency photon detectors~\cite{asdc}.  
A WbLS target~\cite{wbls} allows for the high light yield and correspondingly high energy resolution and low threshold of a scintillator experiment in a directionally sensitive detector.  The admixture of water also increases the attenuation length, thereby allowing the possibility of a very large detector.  
The use of ultra-fast photon detectors would allow separation of the prompt Cherenkov light from delayed scintillation, providing excellent particle ID and background rejection capabilities.  
The ASDC would provide unprecedented sensitivity to solar neutrinos via two channels:
\begin{enumerate}
\item Huge statistics for elastic scattering events at low energy;
\item Potential charged-current detection via isotope loading e.g. $^7$Li~\cite{sigd}.
\end{enumerate}
The unique advantage of the ASDC would be the access to directional information for ES events in a low-threshold detector, providing a powerful tool for background discrimination, in addition to a high-precision spectral measurement using CC events.  The differential cross section for CC neutrino interactions on $^7$Li  is shown in Figure~\ref{f:asdc} in comparison to that for ES interactions.  The distinct shape of the CC differential cross section allows for high precision extraction of the incident neutrino energy by measuring the energy of the recoil electron.  
This could allow both a precision measurement of the $^8$B spectrum, thus providing a sensitive search for new physics, and a measurement of the CNO neutrino flux, including potential sensitivity to the individual components of this flux (Fig.~\ref{f:asdc}).

\subsection{Terrestrial Inputs to the Solar Program}
The solar program relies on a number of critical inputs.  These include nuclear cross section measurements, such as $^3$He($\alpha$, $\gamma$)$^7$Be, $^7$Be(p, $\gamma$)$^8$B, and $^{14}$N(p, $\gamma$)$^{15}$O, from experiments such as LUNA~\cite{luna}.  Terrestrial oscillation parameter measurements from experiments such as KamLAND, Super-Kamiokande, and the proposed GADZOOKS!~\cite{gadzooks} detector can help to constrain global fits, increasing the sensitivity of such fits to potential non-standard interactions and new physics effects.

\section{Conclusions}
The field of solar neutrinos provides a wealth of physics opportunities still to be explored.  
The solar neutrino sector provides the only significant observed matter effect, which causes an additional suppression of the vacuum electron neutrino survival probability at high energies.  This offers 
a unique opportunity to probe the details of the interaction of neutrinos with matter.  The vacuum-matter transition region is also highly sensitive to new physics effects, such as flavor-changing neutral currents.  Hints for such effects have been seen in the global solar neutrino data set,  but to date the precision does not allow for a significant observation~\cite{richie}.  A precision measurement of the CNO neutrino flux would resolve recent ambiguities in the heavy-element composition of the Sun, and could potentially allow determination of the individual primordial abundances of C and N, an important step for nuclear astrophysics and stellar evolution.  A precision measurement of pp neutrinos would allow us to monitor the Sun in close to real time, and to test our understanding of energy loss and generation mechanisms.  

A diverse set of techniques have been used to observe solar neutrinos to date, including radio-chemical experiments, water Cherenkov detectors, and organic liquid scintillator detectors.  The next generation of experiments will need the ability to probe a low-energy regime with high precision in order to address the compelling physics program in this field.  Current experiments will continue to take data, increasing the global solar neutrino data set and offering improved precision in global analyses.  Other multi-purpose detectors, including the upcoming direct dark matter searches, will come online in the near future, offering additional opportunities to explore the solar sector.  The future of this program may lie in multi-purpose detectors, which will be enabled by a new and exciting set of technologies being pursued for next-generation neutrino experiments.  Such experiments could  have the capability to simultaneously resolve many
of the major open questions in neutrino physics, offering breakthroughs both in solar neutrinos and beyond. 

\begin{theacknowledgments}

The author wishes to thank Frank Calaprice, Wick Haxton, and Joshua Klein for many interesting and edifying conversations.  This material is based upon work supported by the Director, Office of Science, of the U.S. Department of Energy under Contract No. DE-AC02-05CH11231; the U.S. Department of Energy, Office of Science, Office of Nuclear Physics, under Award Number DE-SC0010407; the National Science Foundation under Grant No. NSF-PHY-1242509; and the University of California, Berkeley.

\end{theacknowledgments}

\bibliographystyle{aipproc}

\begin{thebibliography}{9}

\bibitem{HRS}W. C. Haxton, R. G. Robertson, A. Serenelli, Ann. Rev. Astron. Astrophys.  {\bf 51} 21-61 (2013).
\bibitem{chlorine} B. T. Cleveland {\it et al.}, ApJ {\bf 496} 505 (1998).
\bibitem{gallex} P. Anselmann {\it et al.}, Nuc. Phys. B (Proc. Suppl.){\bf 31} 117 (1993).
\bibitem{sage}J.N. Abdurashitov \textit{et al.}, Nucl. Phys. Proc. Suppl.{\bf 118} 39 (2003).
\bibitem{gno} M. Altmann {\it et al.}, Phys. Lett. B{\bf 490} 16 (2000).
\bibitem{sk}K. Abe {\it et al.}, Phys. Rev. D{\bf 83} 052010 (2011).
\bibitem{snocc} SNO Collaboration, Phys. Rev. Lett. {\bf 87} 071301 (2001).
\bibitem{Borexpp} G. Bellini {\it et al.}, Nature {\bf 512} 383-386 (2014).
\bibitem{ssm} John N. Bahcall {\it et al.}, ApJ {\bf 621} L85 (2005).
\bibitem{snonc} SNO Collaboration, Phys. Rev. Lett. {\bf 89} 011301 (2002).
\bibitem{kl} KamLAND Collaboration, Phys. Rev. Lett. {\bf 90} 021802 (2003); Phys. Rev. Lett. {\bf 94} 081801 (2005).
\bibitem{kii} K. S. Hirata {\emph et al.}, Phys. Rev. Lett. {\bf 63} 16 (1989). 
\bibitem{wolf} L. Wolfenstein, Phys. Rev. D{\bf 17} 2369 (1978).
\bibitem{ms} S.P. Mikheyev \& A.Y. Smirnov, Sov. J. Nucl. Phys.{\bf 42} 913 (1985).
\bibitem{mavan} M. C. Gonzalez-Garcia, M. Maltoni, Phys. Rept.{\bf 460} 1-129 (2008).
\bibitem{nsi2}A. Friedland, C. Lunardini, C. Pena-Garay, Phys. Lett. B{\bf 594} 347 (2004).
\bibitem{sterile}P. C. de Holanda \& A. Yu. Smirnov, Phys. Rev. D{\bf 83} 113011 (2011).
\bibitem{leta}SNO Collaboration, Phys. Rev. C {\bf 81}:055504 (2010).
\bibitem{skleta}A. Renshaw arXiv:1403.4575 [hep-ex] (2014).
\bibitem{Bor8B}G. Bellini {\it et al.}, Phys. Rev. D{\bf 82} 033006 (2010)
\bibitem{SK8B}M. Smy, Talk at Neutrino 2012, Kyoto (2012)
\bibitem{KL8B}S. Abe {\it et al.}, Phys. Rev. C {\bf 84} 035804 (2011).
\bibitem{richie}R. Bonventre {\it et al.}, Phys. Rev. D {\bf 88}:053010 (2013).
\bibitem{richie2}R. Bonventre {\it Personal communication}.
\bibitem{asplund}M. Asplund {\it et al.}, Ann. Rev. Astron. Astrophys.{\bf 47} 481 (2009).
\bibitem{met1}A. Serenelli {\it et al.}, ApJ{\bf 705} 2 (2009).
\bibitem{met2}A. Serenelli, W. C. Haxton, C. Pena-Garay,  ApJ{\bf 743} 1 (2011).
\bibitem{skdn} A. Renshaw {\emph et al.}, Phys. Rev. Lett. {\bf 112} 091805 (2014).
\bibitem{hk} Hyper-Kamiokande Working Group arXiv:1309.0184 [hep-ex] (2013).
\bibitem{Borex7be} G. Bellini {\emph et al.}, Phys. Rev. Lett. {\bf 107} 141302 (2011).
\bibitem{Borexpep}G. Bellini {\it et al.},  Phys. Rev. Lett. {\bf 108} 051302 (2012).
\bibitem{lena} M. Wurm  {\it et al.}, arXiv:1104.5620 [astro-ph.IM] (2011).
\bibitem{JUNO} JUNO Collaboration \url{http://english.ihep.cas.cn/rs/fs/juno0815/}

\bibitem{lbne}C. Adams, \emph{et al.}, arXiv: 1307.7335 (2014).
\bibitem{clean} M.G. Boulay, A. Hime, and J. Lidgard arXiv: nucl-ex/0410025 (2004); 
D. N. McKinsey, K. J. Coakley Astropart. Phys. {\bf 22} 355 (2005).
\bibitem{lz} D. C. Malling {\emph et al.}, arXiv:1110.0103 [astro-ph.IM] (2011).
\bibitem{xmass} K. Abe {\emph et al.}, Nucl. Instr. Meth. A {\bf 716} 78 (2013).
\bibitem{asdc} The ASDC Working Group arXiv: 1409.5864 [physics.ins-det] (2014).

\bibitem{lens}C. Grieb \& R. Raghavan, Phys. Rev. Lett.{\bf 98} 141102 (2007).
\bibitem{rcli}A. Kopylov \& V. Petukhov, arXiv:hep-ph/0301016, arXiv:hep-ph/0306148,  	arXiv:hep-ph/0308004 (2003); 
A. Kopylov {\it et al.}, {\it Phys. Atom. Nucl.}{\bf 67} 1182 (2004); 
A. Kopylov {\it et al.}, {\it Phys. Atom. Nucl.}{\bf 69} 1829 (2006); 
A. Kopylov \& V. Petukhov, {\it JCAP}{\bf 10} 007 (2008); 
A. Kopylov {\it et al.}, arXiv:0910.3889 [nucl-ex] (2009).
\bibitem{sigd}W. Haxon,  Phys. Rev. Lett.{\bf 76} 10 (1996).



\bibitem{wbls}M. Yeh {\it et al.},Nucl. Inst. \& Meth. A{\bf 660} 51 (2011).
\bibitem{luna} LUNA Collaboration, Nucl. Phys. Sec. {\bf A814} 144-158 (2008).
\bibitem{gadzooks}  J. F. Beacom and M. R. Vagins,  Phys.Rev. Lett. {\bf 93} 171101 (2004).


\end{thebibliography}

\end{document}